\documentstyle[12pt]{article}
\begin{document}
\def\thefootnote{\fnsymbol{footnote}}
\begin{flushright}
KANAZAWA-01-03  \\ 
April, 2001
\end{flushright}
\vspace*{2cm}
\begin{center}
{\LARGE\bf The origin of quark and lepton mixings}\\
\vspace{1 cm}
{\Large  Daijiro Suematsu}
\footnote[1]{e-mail:suematsu@hep.s.kanazawa-u.ac.jp}
\vspace {0.7cm}\\
$^\ast${\it Institute for Theoretical Physics, Kanazawa University,\\
        Kanazawa 920-1192, JAPAN}
\end{center}
\vspace{2cm}
{\Large\bf Abstract}\\  
We propose a model for the mass matrices of quarks and leptons 
based on two Abelian flavor symmetries. One is assumed to be broken
at a high energy region near the Planck scale. It is used for the
Froggatt-Nielsen mechanism in both quark and charged lepton sectors. 
Another one remains unbroken to a multi-TeV region. The mixings among 
neutrinos and gauginos including the one of the new Abelian symmetry
generate non-zero masses and mixings among neutrinos. 
A bi-maximal scheme for the neutrino oscillation can be realized 
together with suitable masses and CKM-mixings in the quark sector.
FCNC constraints on this flavor dependent Abelian symmetry seems to
be evaded.
\newpage
\setcounter{footnote}{0}
\def\thefootnote{\arabic{footnote}}
\noindent
{\Large\bf 1.~Introduction}

The origin of flavor mixings of quarks and leptons is one of the
most important problems beyond the standard model (SM).
Recently the existence of non-trivial lepton mixings has been strongly
suggested through the atmospheric and solar neutrino observations 
whose results can be explained by assuming the neutrino oscillations 
\cite{oscil1,oscil2,sk}. 
The predicted flavor mixing is much bigger than the one of the quark
sector. The explanation of this feature is a challenge to the
grand unified theory (GUT) and a lot of works have been done
by now \cite{gut,af}. In many models the flavor mixing in both sectors is 
considered to be controlled by the Froggatt-Nielsen mechanism 
\cite{fn}\footnote{There are many works in which the Abelian flavor 
symmetry is discussed. As examples, see
\cite{fnap} and references therein.} and the smallness of the neutrino 
mass is explained by the celebrated seesaw mechanism \cite{seesaw}.
In such scenarios the origin of the flavor mixings is eventually
related to the physics at high energies such as the Planck scale
and an intermediate scale.

In this paper we propose an alternative possibility based on a
different origin of
the flavor mixings of neutrinos in a supersymmetric model with an
extended gauge structure. 
In our scenario the flavor mixings in the quark and charged lepton 
sectors are considered to have its origin at the high energy region 
due to the usual Froggatt-Nielsen mechanism. 
On the other hand, the flavor mixings in the neutrino sector are assumed to
come from their mixings with the extended gaugino sector by an extra
U(1) gauge symmetry \cite{gaugino}.
The mixings are induced by the R-parity violation \cite{rparity1,rparity2}. 
Its origin might be considered to be related to the physics concerning
to an effective supersymmetry breaking at a TeV region. 
One of the interesting points of the model is that
the large mixing angle MSW solution for the solar neutrino
problem can be consistently accommodated together with the small
flavor mixings and the qualitatively favorable mass eigenvalues 
in the quark sector. 
The scenario could be consistent with the gauge coupling unification 
since it has an SU(5) GUT structure when we switch off the low energy
extra U(1) gauge interaction which plays a part of flavor symmetry
together with the Froggatt-Nielsen type U(1)-symmetry.  
For convenience, we will use the SU(5) representations to classify 
the fields in the following discussion.

In the next section we define the flavor symmetry of the model.
In section 3 the mass matrices in the quark and charged lepton 
sectors are discussed. FCNC constraints are also examined for the 
non-universal couplings of an extra neutral gauge boson here. 
The neutrino mass matrix is studied in section 4.
We show that a bi-maximal mixing is derived in this model. We also
discuss the $R$-parity violation here. Section 5 is devoted to the summary.
\vspace*{3mm}

\noindent
{\Large\bf 2.~Abelian flavor symmetry}

We consider a model with two Abelian flavor symmetries
U(1)$_F\times$U(1)$_X$.
The U(1)$_F$ is considered to be broken near the Planck scale
and used to generate the mass hierarchy and the flavor mixing 
through the Froggatt-Nielsen mechanism. It may be considered to
be an anomalous U(1) symmetry.
On the other hand, the U(1)$_X$ is assumed to remain unbroken to 
the TeV region\footnote{Additional U(1)-symmetries are known 
to appear very often in the heterotic superstring models and 
it can play some useful roles in the supersymmetric models \cite{z00}.}.
We assume that this low energy extra U(1)$_X$ gauge field 
has flavor diagonal but non-universal couplings. 
Different charges of U(1)$_X$ are assigned to the ${\bf 5^\ast}$
fields belonging to the different generations\footnote{This 
kind of charge assignment of U(1)$_X$ has often been discussed in the different
context. For example, it has been assumed to explain the small neutrino 
mass and the proton stability in \cite{sue2}.}.
Since its breaking scale is in the TeV region, we cannot use this 
symmetry for the Froggatt-Nielsen mechanism to induce the 
hierarchical structure of quark mass matrices since its breaking scale 
is too small as compared to the Planck scale.
However, since a part of the contents of the minimal supersymmetric SM
(MSSM) is assumed to have its 
charge in a generation dependent way, it can generate an additional 
non-trivial texture in the mass matrices.

We adopt the following charge assignments of
U(1)$_{F}\times$U(1)$_X$ to the chiral superfields of quarks and
leptons:  
\begin{equation}\begin{array}{cccc}
{\bf 10}_f\equiv (q, u^c, e^c)_f &:& (3, 2, 0), &  
 (\alpha, \alpha, \alpha), \\ 
{\bf 5^\ast}_f \equiv (d^c, \ell)_f &:& (1, 0, 0), & 
(q_1, q_1, q_2).\end{array}
\label{eqa}
\end{equation}
where $f$=1-3 and the numbers in the parentheses stand for the charges 
for each generation. We need no right-handed neutrino since in the present 
scenario neutrino masses are assumed to be generated through the
mixings with gauginos. 
In general, the introduction of U(1)$_X$ to the MSSM requires additional
chiral superfields to cancel the gauge anomaly which causes the non-trivial
constraint on the charge assignment. 
Since it is assumed to be radiatively broken at the TeV region,
we need at least a new SM singlet chiral superfield whose scalar 
component causes 
the spontaneous breaking of U(1)$_X$ by its vacuum expectation value (VEV). 
Taking account of these, as the Higgs chiral superfield sector 
we consider the following contents:
\begin{equation}
\begin{array}{cccc}
{\bf 5}^a \equiv (D; H_2)^a &:& ((0,0);(0,0)),&  
 ((x,z);(p,r)), \\ 
{\bf 5^\ast}^a \equiv (\bar D; H_1)^a &:& ((0,0);(0,0)),&  
 ((y,w);(q,s)), \\ 
{\bf 1}_0 \equiv S_0  &:& (-1),& (0), 
\\ 
{\bf 1}_i \equiv S_i  &:&  (0,\cdots, 0), & (Q_1,\cdots, Q_6),
\end{array}
\label{eqb}
\end{equation} 
where $a$=1-2 and $i$=1-6.
From the charge assignment of U(1)$_X$ for ${\bf 5}^a$ and 
${\bf 5}^{\ast a}$ we find that the SU(5) symmetry is explicitly 
broken unless U(1)$_X$ 
is switched off. Then the SU(5) has only a meaning as a classification group 
in the model.
The choice of additional chiral superfields and their charges should guarantee 
the SM gauge coupling unification,
the proton stability and the anomaly cancellation of U(1)$_X$.
The coupling unification of the SM gauge group is expected to be 
satisfied as a result of the SU(5) structure\footnote{
The coupling unification requires an additional couple of
SU(2)$_L\times$U(1)$_Y$ vector like fields such as $H_1$ and
$H_2$. However, since they can be trivial under the U(1)$_X$ and obtain
masses, for example, due to the Giudice and Masiero 
mechanism \cite{gm}, they can be irrelevant to our discussion.}.
Moreover, the U(1)$_X$ can prohibit the couplings between quarks and 
extra colored fields $D^a$ and $\bar D^a$ and guarantees 
the proton longevity.

Due to other phenomenological requirements some conditions should be
imposed on the U(1)$_X$. 
It should allow the presence of necessary Yukawa couplings to generate
the masses of quarks and leptons. For this requirement we impose the
conditions
\begin{equation}
2\alpha+p=0, \quad \alpha +q_1+q=0, \quad \alpha +q_2+s=0,
\label{eqc}
\end{equation}
and then we have the following Yukawa couplings:
\begin{eqnarray}
W_{\rm Yukawa}&=&\sum_{f,f^\prime=1}^3y^u_{_{ff^\prime}}u^c_{_f}
H_2^1q_{_{f^\prime}}
\nonumber \\
&+&\sum_{f^\prime=1}^3\left\{\sum_{f=1,2}\left(
y^d_{_{ff^\prime}}d^c_{_f}H_1^1q_{_{f^\prime}}
+y^e_{_{ff^\prime}}e^c_{_f}H_1^1\ell_{_{f^\prime}}\right)
+y^d_{_{3f^\prime}}d^c_3H_1^2q_{_{f^\prime}}
+y^e_{_{3f^\prime}}e^c_3H_1^2\ell_{_{f^\prime}}
\right\}.
\label{eqd}
\end{eqnarray}
We also require the following conditions on the Higgs chiral superfield sector:
\begin{eqnarray}
&&p+q+Q_1=0, \quad p+ s+Q_2=0, \quad x+y+ Q_3=0, \quad Q_1+Q_2+Q_3=0, 
\nonumber \\
&&r+q+Q_4=0, \quad r+ s+Q_5=0, \quad z+w+ Q_6=0, \quad Q_4+Q_5+Q_6=0. 
\label{eqe}
\end{eqnarray}
We assume that the scalar components of the SM singlet fields 
$S_3$ and $S_6$ get VEVs at the TeV
region radiatively through the couplings with extra colored fields $D^a$
and $\bar D^a$ and then the U(1)$_X$ gauge field becomes massive \cite{z00}. 
Moreover, the last conditions in each line of (\ref{eqe}) allow the trilinear
couplings $S_1S_2S_3$ and $S_4S_5S_6$ in the superpotential. 
These trilinear couplings are accompanied with scalar trilinear
couplings which break the supersymmetry softly
and they can generally induce the VEVs to other singlet fields $S_i$.
As a result all of the extra colored fields $D^a$ and 
$\bar D^a$ and the doublet Higgs fields become massive at that scale
through the couplings with $S_i$. 
The mixings of the doublet Higgs fields in the superpotential 
can be written as
\begin{equation}
(H_2^1, H_2^2)\left(\begin{array}{cc}
\kappa_1\langle S_1\rangle & \kappa_2\langle S_2\rangle \\
\kappa_4\langle S_4\rangle & \kappa_5\langle S_5\rangle \\
\end{array}\right)\left(\begin{array}{c}H_1^1 \\ H_1^2 \\
\end{array}\right).
\label{eqf}
\end{equation}  
If ${\kappa_5\langle S_5\rangle\over \kappa_4\langle S_4\rangle}$
is equal to $-{\kappa_1\langle S_1\rangle\over 
\kappa_2\langle S_2\rangle}$, the eigenstates of the mixing matrix 
(\ref{eqf}) can be
identified as $(H_2^1, H_1^{\ell}\equiv\sin\zeta H_1^1+\cos\zeta H_1^2)$
and $(H_2^2, H_1^{h}\equiv -\cos\zeta H_1^1+\sin\zeta H_1^2)$ where $\tan\zeta=
{\kappa_1\langle S_1\rangle\over \kappa_2\langle S_2\rangle}$.
Since $H_2^1$ has a coupling with top quark as shown in eq.~(\ref{eqd})
and a mixing with $H_1^\ell$ through eq.~(\ref{eqf}),
only the former set of Higgs fields is expected to get the VEVs and it
works as the usual Higgs fields. In the following discussion we
take it as a basic assumption since the mixing in the $H_1^a$ sector can 
play an important role to derive the MNS matrix due to the effect on the 
charged lepton mass matrix. 
Anomaly free conditions for U(1)$_X$ can impose an addtional constraint.
If we require SU(3)$_C\times$SU(2)$_L\times$U(1)$_Y\times$U(1)$_X$ 
to be anomaly free under the conditions (\ref{eqc}) and (\ref{eqe}), 
the U(1)$_X$ charges $q_1$ and $q_2$ will be constrained into the 
restricted region. 
The details are discussed in an appendix and
we show a few examples of the solutions in Fig.~1.  
\vspace*{3mm}

\noindent
{\Large\bf 3.~Mass and mixing of quarks and charged leptons}

The U(1)$_F$ symmetry controls the flavor mixing structure by regulating 
the number of field $S_0$ contained in each non-renormalizable term
through the so-called Froggatt-Nielsen mechanism.
As its result the effective Yukawa couplings $y_{_{ff^\prime}}$ in 
eq.~(\ref{eqd}) have the hierarchical structure. 
In fact, if the singlet field $S_0$ gets the VEV $\langle S_0\rangle$, 
the suppression factor for the Yukawa 
couplings could appear as the power of $\lambda=
{\langle S_0\rangle\over M_{\rm pl}}$. Here $M_{\rm pl}$ is the Planck scale.
Using the U(1)$_F$ charges introduced above, we can obtain the mass matrices
of quarks and charged leptons in the usual way.
However, there is additional structure coming from the U(1)$_X$
constraints which are realized by the condition (\ref{eqc}) and 
a composition of the doublet Higgs field $H_1^{\ell}$.
Taking them into account we can write the mass matrices for 
the quarks and the charged leptons as follows:
\begin{eqnarray}
&&M_u\sim \left(\begin{array}{ccc}\lambda^6 &\lambda^5 &\lambda^3 \\
\lambda^5 &\lambda^4 &\lambda^2 \\ \lambda^3 &\lambda^2 & 1 \\
\end{array}\right)\langle H_2^1\rangle, \quad
M_d\sim \left(\begin{array}{ccc}\lambda^{4}\sin\zeta &\lambda^{3}\sin\zeta 
& \lambda\sin\zeta \\
\lambda^3\sin\zeta &\lambda^2\sin\zeta & \sin\zeta \\ \lambda^3\cos\zeta 
&\lambda^2\cos\zeta & \cos\zeta \\
\end{array}\right)\langle H_1^\ell\rangle, 
\label{eqg}
\end{eqnarray} 
where the above mass matrices are written in the basis of 
$\bar\psi_Rm_D\psi_L$. We do not consider the CP phases here. 
In eq.~(\ref{eqg}) we abbreviate the order one coupling constants by 
using the similarity symbol. In the quark sector the mass
eigenvalues and the elements of the CKM matrix can be found after some
inspection as
\begin{eqnarray}
&& m_u : m_c : m_t = \lambda^6 : \lambda^4 : 1, \nonumber \\
&& m_d : m_s : m_b = \lambda^4\sin\zeta : \lambda^2\cos\zeta : \cos\zeta, 
\nonumber \\
&& V_{us} \sim \lambda, \qquad V_{ub}\sim \lambda^3, 
\qquad V_{cb}\sim \lambda^2. 
\label{eqh}
\end{eqnarray}
If we require that the order one couplings should be in a range
$(\sqrt\lambda, {1\over\sqrt\lambda})$, the lower bound of
$\tan\beta\equiv \langle H_2^1\rangle/\langle H_1^\ell\rangle$ is
estimated as $\tan\beta~{^>_\sim}~40\lambda\cos\zeta$.
On the charged lepton sector we can know the mass eigenvalues by 
noting that the SU(5) relation such as $M_e^T=M_d$ is also satisfied 
in the present model. The ratio of mass 
eigenvalues is the same as the one of the down quark sector and then
\begin{equation}
m_e : m_\mu : m_\tau = \lambda^4\sin\zeta : \lambda^2\cos\zeta : \cos\zeta.
\label{eqi} 
\end{equation}
The result has some different features from
the ones presented in ref.~\cite{af} in the down quark and charged 
lepton sectors. 
It comes from the charge assignment for ${\bf 5}^\ast$ and the
composition of the Higgs field $H_1^\ell$. 
If we assume $\lambda\sim 0.22$, these results seem to 
describe nicely the experimental data for the mass eigenvalues and the
CKM-mixing angles as far as $\cos\zeta\sim\sin\zeta$ is satisfied, 
except for $m_u$ and $m_e$ which are predicted to be somehow larger.
Also in our framework we cannot overcome this common defect with the
scheme based on a kind of U(1)$_F$ symmetry. 
The value of $\sin\zeta$ is, in principle, determined by the scalar
potential of the singlet fields $S_i$ which is briefly discussed below 
(\ref{eqe}). From its structure the above value of $\sin\zeta$ seems to be 
expected to be realized without unnatural tunings of parameters.   
We define the diagonalization matrix $\tilde U$ of the charged 
lepton mass matrix in a basis that $\tilde U^\dagger M_e^\dagger
M_e\tilde U$ is diagonal. Then $\tilde U$ can be approximately written as
\begin{equation}
\tilde U=\left(\begin{array}{ccc}
1 & 0 & \lambda\sin\zeta\\
-\lambda\sin^2\zeta& \cos\zeta & \sin\zeta \\
-\lambda\sin\zeta\cos\zeta & -\sin\zeta & \cos\zeta \\
\end{array}\right),
\label{eqjj}
\end{equation}  
where the CP phase in the charged lepton sector is assumed to be zero
although there can be some sources for it. 

The breaking scale of U(1)$_X$ is assumed to be in the TeV region and 
then we have a rather light $Z^\prime$ which can impose constraints on 
the model. 
The non-universal couplings of $Z^\prime$ with 
${\bf 5}^\ast$ may induce large flavor changing neutral currents (FCNC). 
The detailed analysis for such an issue has been done in \cite{fcnc} 
and we can apply the discussion to the present model. 
In the model the $Z^\prime$ interaction term relevant to 
the non-universal couplings of 
${\bf 5}^\ast$ can be written in the mass eigenstates as,
\begin{eqnarray}
&&{\cal L}_{Z^\prime}=-g_1\left[{g_X\over g_1}\cos\xi J_{(2)}^\mu
-\sin\xi J_{(1)}^\mu\right] Z_\mu^{(2)}, \nonumber \\
&&J_{(2)}^\mu=\sum_{ij}\left[\bar\nu_{L_i}B_{ij}^{\nu_L}\nu_{L_j}
+\bar\ell_{L_i}B_{ij}^{\ell_L}\ell_{L_j}
+\bar d_{R_i}B_{ij}^{\ell_R}d_{R_j}\right], \nonumber \\ 
&&B_{ij}^\psi= V^{\psi\dagger}~ {\rm diag}(q_1, q_1, q_2)~V^\psi,
\label{eqkk}
\end{eqnarray}
where $J_{(1)}^\mu$ is the SM weak neutral current and $\xi$ 
is a $Z$-$Z^\prime$ mixing angle.  $V^\psi$ is a
unitary matrix used to diagonalize the mass matrix of $\psi$.
In the present model $V^{d_R}=V^{\ell_L}$ is satisfied because of 
the SU(5) relation and they are represented by eq.~(\ref{eqjj}).
Thus the relevant $B_{ij}^\psi$ can be estimated as
\begin{eqnarray}
&&\vert B_{12}^{\ell_L}\vert^2=\vert B_{12}^{d_R}\vert^2=
(q_1-q_2)^2\lambda^2\sin^4\zeta\cos^2\zeta, \nonumber \\
&&\vert B_{13}^{\ell_L}\vert^2=\vert B_{13}^{d_R}\vert^2=
(q_1-q_2)^2\lambda^2\cos^4\zeta\sin^2\zeta, \nonumber \\
&&\vert B_{23}^{\ell_L}\vert^2=\vert B_{23}^{d_R}\vert^2=
(q_1-q_2)^2\cos^2\zeta\sin^2\zeta.
\label{eqk}
\end{eqnarray} 
On the other hand, the experimental constraints on these values are also
estimated in \cite{fcnc}. The relevant constraints to the present model 
can be summarized as follows\footnote{We do not consider the 
CP violating effect here.}. The coherent $\mu$-$e$ conversion and the 
decays $\tau\rightarrow 3e, 3\mu$ require
\begin{equation}
w^2\left(\vert B_{12}^{\ell_L}\vert^2 \right)
< 4\times 10^{-14}, \qquad
w^2\left(\vert B_{13}^{\ell_L}\vert^2 \right)
< 2\times 10^{-5}, \qquad
w^2\left(\vert B_{23}^{\ell_L}\vert^2  \right)
< 10^{-5}, 
\label{eql}
\end{equation}
and from the lepton flavor violating meson decays such as
$K_L\rightarrow\mu^\pm e^\mp$ we obtain
\begin{equation}
y^2\vert B_{12}^{\ell_L}\vert^2\vert B_{12}^{d_R}\vert^2<10^{-14}.
\end{equation}
The lepton flavor conserving meson decays such as
$K_L\rightarrow\mu^+\mu^-$ impose
\begin{equation}
w^2\vert{\rm Re}\left[ B_{12}^{d_R}\right]\vert^2< 3\times 10^{-11},
\end{equation}
and decays of $B^0$ into $\mu^+\mu^-$ and $\pi^0\mu^+\mu^-$ give
\begin{equation}
w^2\vert B_{13}^{d_R}\vert^2< 10^{-5}, \qquad
w^2\vert B_{23}^{d_R}\vert^2< 3\times 10^{-6}.
\end{equation}
In addition, from the experimental results on meson mass splittings 
we know the conditions
\begin{equation}
y\vert {\rm Re}\left[\left(B_{12}^{d_R}\right)^2\right]\vert< 10^{-8}, \quad
y\vert {\rm Re}\left[\left(B_{13}^{d_R}\right)^2\right]\vert<6\times 
10^{-8}, \quad
y\vert {\rm Re}\left[\left(B_{23}^{d_R}\right)^2\right]\vert<2\times 
10^{-8}
\end{equation}
should be satisfied. In these conditions we use the definition 
\begin{equation}
w={g_X\over g_1}(\rho_1-\rho_2)\sin\xi\cos\xi, \qquad 
y=\left({g_X\over g_1}\right)^2(\rho_1\sin^2\xi+\rho_2\cos^2\xi),  
\end{equation}
where $\rho_i ={M_W^2\over M_{Z_i}^2\cos^2\theta_W}$.
If we apply the above value of $\lambda$ and 
$\sin\zeta\sim{1\over\sqrt 2}$ to eq.~(\ref{eqk}), 
the most stringent constraints on $w$
and $y$ are obtained as\footnote{In this
discussion we assume that the U(1)$_X$-charge is normalized in the same
way as U(1)$_Y$.} 
\begin{equation}
w^2(q_1-q_2)^2 < 7\times 10^{-12}, \qquad 
y(q_1-q_2)^2 < 7\times 10^{-12}.
\label{eqll}
\end{equation}
If $\vert q_1 -q_2\vert$ takes the value of order one,
$w$ and $y$ should be smaller than $10^{-6}$.
They require $\sin\xi < 10^{-6}$ and $M_{Z_2}>100$~TeV.
The value of charges $q_1$ and $q_2$ will be discussed from the view
point of the explanation of the solar and atmospheric neutrino 
problems in the next section. 
\vspace*{3mm}

\noindent
{\Large\bf 4.~Neutrino mass and mixing}

On the neutrino mass and mixing we adopt the scenario proposed in
\cite{gaugino}. 
Using the U(1)$_X$ charges defined in (\ref{eqa}), the coupling between 
the neutrinos and the U(1)$_X$ gaugino is given by
$i\sqrt 2g_{_X}\sum_\alpha q_\alpha\left(\tilde\nu_\alpha^\ast
\lambda_X\nu_\alpha -\bar\lambda_X\bar\nu_\alpha\tilde\nu_\alpha\right)$.
We do not consider the kinetic term mixing between the U(1)-gauginos
\cite{sue1}. If we take this effect into account, 
off-diagonal elements appear in the gaugino mass matrix.
The U(1)$_X$ can plays an crucial role in the generation of
non-zero neutrino masses due to the above mentioned interaction since
$\nu_{\tau}$ has a charge $q_{2}$ different from other neutrinos $\nu_e$ and 
$\nu_{\mu}$ whose charges are defined by $q_1$. 
If sneutrinos get the VEV $u$ due to the R-parity violation, 
the mixing among neutrinos and gauginos appear as ${\cal L}_{\rm
mass}=-{1\over 2}({\cal N}^T{\cal M}{\cal N}+ {\rm h.c.})$ and
\begin{equation}
{\cal M}=\left(\begin{array}{cc}0 &m^T \\ m & M \\ \end{array} \right),
\qquad
m=\left(\begin{array}{ccc}
a_2 & a_1 & b \\ a_2 & a_1 & b \\ a_2 & a_1 & c \\
\end{array}\right), 
\qquad
M=\left(\begin{array}{ccc}
M_2 & 0 & 0 \\ 0 & M_1 & 0 \\ 0 & 0 & M_X \\
\end{array}\right),
\label{eqm}
\end{equation}
where ${\cal N}^T=(\nu_\alpha, -i\lambda_{W_3}, i\lambda_Y,
-i\lambda_X)$. We use the definition such as 
$a_{\ell}={g_{\ell}\over\sqrt 2}u$, $b=\sqrt 2g_{_X}q_1u$ 
and $c=\sqrt 2g_{_X}q_2 u$.  
If we assume $u$ is much smaller than the gaugino masses $M_A$,
we can obtain the light neutrino mass matrix from it by
using the generalized seesaw formula. It can be written as
\begin{equation}
M_\nu = m^T M^{-1}m =\left(\begin{array}{ccc}
m_0+\epsilon^2 & m_0+\epsilon^2 & m_0+\epsilon\delta \\
m_0+\epsilon^2 & m_0+\epsilon^2 & m_0+\epsilon\delta \\
m_0+\epsilon\delta & m_0+\epsilon\delta & m_0+\delta^2 \\
\end{array}\right),
\label{eqn}
\end{equation}
where $m_0,\epsilon$ and $\delta$ are defined by 
\begin{equation}
m_0={g_2^2u^2\over 2M_2}+{g_1^2u^2\over 2M_1}, \qquad 
\epsilon={\sqrt 2g_{_X}q_1u\over\sqrt{M_X}}, \qquad
\delta={\sqrt 2g_{_X}q_2u\over\sqrt{M_X}}.
\label{eqo}   
\end{equation}
If the neutrinos have the same U(1)$_X$ charge, we find that 
there is only one nonzero mass eigenvalue as in the usual R-parity
violating scenario \cite{rparity1,rparity2}.
The interesting aspect of this mass matrix is that it is defined only by the
 gaugino mass $M_A~(A=\ell, X)$, the gauge couplings $g_{_A}$, the 
U(1)$_X$-charges $q_\alpha$ and the VEV $u$ of sneutrinos. 
We define the mass eigenstates $\nu_i$ by 
$\nu_\alpha= U_{\alpha i}\nu_i$. 
A mass eigenvalue $m_2$ is zero and non-zero mass eigenvalues are
represented as
\begin{equation}
m_{1,3}
={1\over 2}\left\{(3m_0+2\epsilon^2+\delta^2)
\pm\sqrt{(m_0+2\epsilon^2-\delta^2)^2+8(m_0+\epsilon\delta)^2}\right\}.
\label{eqr}
\end{equation}
Here we should note that the gaugino mass can have a CP phase which
depends on the supersymmetry breaking mechanism. Because of it we can
consider both possibilities of $\vert m_1\vert <\vert m_3\vert$ and
$\vert m_1\vert >\vert m_3\vert$ depending on a sign of $M_A$.
The diagonalization of the matrix (\ref{eqn}) gives
\begin{equation}
U=\left(\begin{array}{ccc}
{1\over\sqrt 2}\cos\theta & -{1\over\sqrt 2} & -{1\over\sqrt 2}\sin\theta \\
{1\over\sqrt 2}\cos\theta & {1 \over\sqrt 2} & -{1\over\sqrt 2}\sin\theta  \\
\sin\theta & 0 & \cos\theta \\
\end{array}\right),
\label{eqp}
\end{equation}
where one of the mixing angles $\sin\theta$ is defined as
\begin{equation}
\sin^22\theta ={8(m_0+\epsilon\delta)^2\over (m_0+2\epsilon^2-\delta^2)^2
+8(m_0+\epsilon\delta)^2}.
\label{eqq}
\end{equation}
If we remind that the MNS-matrix which controls the neutrino 
oscillation phenomena is
defined as $V^{\rm (MNS)}=U^T\tilde U$, we find that it can be written 
by using eqs.~(\ref{eqjj}) and (\ref{eqp}) as
\begin{equation}
\footnotesize
\left(\begin{array}{cccc}
{\cos\theta\over \sqrt 2}(1-\lambda\sin^2\zeta)
-\lambda\sin\theta\sin\zeta\cos\zeta & 
-{1\over\sqrt 2}(1+\lambda\sin^2\zeta) &
-{\sin\theta\over \sqrt 2}(1-\lambda\sin^2\zeta)
-\lambda\cos\theta\sin\zeta\cos\zeta  \\
{\cos\theta\over \sqrt 2}\cos\zeta-\sin\theta\sin\zeta 
&{\cos\zeta\over \sqrt 2} &
-{\sin\theta\over \sqrt 2}\cos\zeta-\cos\theta\sin\zeta \\
{\cos\theta\over\sqrt 2}(1+\lambda)\sin\zeta +\sin\theta\cos\zeta   
& {1\over \sqrt 2}(1-\lambda)\sin\zeta  &
-{\sin\theta\over\sqrt 2}(1+\lambda)\sin\zeta+\cos\theta\cos\zeta \\
\end{array}\right).
\normalsize
\label{eqt}
\end{equation}

Now we study the oscillation phenomena in the present model.
It is well-known that the transition probability due to the
neutrino oscillation $\nu_\alpha\rightarrow \nu_\beta$ after 
the flight length $L$ is written by using the matrix elements of 
(\ref{eqt}) as
\begin{equation}
{\cal P}_{\nu_\alpha\rightarrow \nu_\beta}(L)
=\delta_{\alpha\beta}
-4\sum_{i>j}V^{\rm (MNS)}_{\alpha i}V^{\rm (MNS)}_{\beta i}
V^{\rm (MNS)}_{\alpha j}V^{\rm (MNS)}_{\beta j}
\sin^2\left({\Delta m_{ij}^2\over 4E}L\right),
\label{equ}
\end{equation}
where $\Delta m_{ij}^2=\vert m_i^2-m_j^2\vert$.
If we remember $m_2=0$, we find that $\Delta m^2_{12}$ or 
$\Delta m^2_{23}$ should be relevant to the atmospheric neutrino
problem. Unless both $\vert\cos\theta\vert\sim 1$ and 
$\vert\sin\zeta\vert\sim\vert\cos\zeta\vert$ are 
not satisfied, the relevant amplitude 
$-4\sum_{j=1,3}V^{\rm (MNS)}_{\mu 2}V^{\rm (MNS)}_{\tau 2}V^{\rm (MNS)}_{\mu j}
V^{\rm (MNS)}_{\tau j}$ to $\nu_\mu\rightarrow\nu_\tau$ is too small 
to explain the atmospheric neutrino deficit.
Thus we assume $\cos\theta\sim 1$ and 
$\sin\zeta\sim\cos\zeta\sim{1\over\sqrt 2}$ here\footnote{The former
condition requires the restricted values for $q_1$ and $q_2$ as is seen
below. The latter one seems to be naturally satisfied as commented before.}.
Under the assumption we obtain
\begin{equation}
V^{\rm (MNS)}\sim \left(\begin{array}{ccc}
{1\over\sqrt 2} & -{1\over\sqrt 2} & -{\lambda\over 2}\\
{1\over 2} & {1\over 2} & -{1\over\sqrt 2}\\
{1\over 2}  & {1\over 2}  &  {1\over \sqrt 2}\\
\end{array}\right).
\label{eqv}
\end{equation}
This MNS-matrix is just the one representing the bi-maximal mixing
\cite{bimaxim,bimaxim2,bimaxim3}.
In our model the large mixing between $\nu_e$ and $\nu_\mu$ has 
an origin in the neutrino sector and is related to the value of
$\cos\theta$. On the other hand, the large mixing between $\nu_\mu$ and
$\nu_\tau$ comes from the charged lepton sector and is relevant to
the value of $\cos\zeta$. This feature makes it possible to meet the MNS 
matrix with the best fit value of the large mixing angle MSW solution
(LMA) to the Super-Kamiokande data of the solar neutrino.
It is a different situation from other bi-maximal mixing models 
based on the Zee model \cite{zee}.
On the squared mass diffrences the atmospheric neutrino deficit 
should be relevant to both $\Delta m^2_{12}$ and 
$\Delta m^2_{23}$. On the other hand, the solar neutrino deficit should be
related to $\nu_e\rightarrow\nu_\mu$ with $\Delta m^2_{12}$ and
$\nu_e\rightarrow\nu_\tau$ with $\Delta m^2_{12}$. It means that the
normal hierarchy $\vert m_2\vert ~{^<_\sim}~\vert m_1\vert \ll \vert 
m_3\vert$ should be satisfied and then $M_A$ is required to have 
a negative sign in eq.~(\ref{eqm}). 
The CHOOZ experiment \cite{chooz} constrains a component $V^{\rm (MNS)}_{e3}$ 
of the MNS-mixing matrix since the amplitude of the contribution to 
$\nu_e \rightarrow \nu_x$ with $\Delta m_{23}^2$ or $\Delta m_{13}^2$
always contains it. The model escapes the constraint since
$\vert V^{\rm (MNS)}_{e3}\vert$ takes 
${\lambda\over 2}$ which satisfies its upper bound $0.16$ \cite{ue3}.  
The value of $V^{\rm (MNS)}_{e3}$ is a prediction of our model.
In the model the effective mass parameter which appears in a
formula of the rate of neutrinoless double $\beta$-decay \cite{bb0}
is estimated as
\begin{equation}
\vert m_{ee}\vert=\left\vert \sum_j\vert V^{\rm MNS}_{ej}
\vert^2e^{i\phi_j}m_j\right\vert
\sim {1\over 2}\vert m_1\vert+ {\lambda^2\over 4}\vert m_3\vert .
\end{equation}
Since $\vert m_1\vert=\sqrt{\Delta m^2_{\rm solar}}$ and 
$\vert m_3\vert=\sqrt{\Delta m^2_{\rm atm}}$, 
it seems to be difficult that $m_{ee}$ will be within the reach 
in near future.

Whether the above values of $\cos\theta$ and mass
eigenvalues $m_i$ can be consistently realized is a crucial point for
the model.
We can study it by using the parameters $M_A$, $g_{_A}$, $q_\alpha$ and $u$.
In the usual soft supersymmetry breaking scenario the gaugino mass
is considered to be universally produced as $M_0$ at the unification scale. 
Its low energy value is determined by the 
renormalization group equations (RGEs).
If we use the one-loop RGEs, the gaugino mass at a scale $\mu$ 
can be expressed as
\begin{equation}
M_2(\mu) = {M_0\over g_U^2}g_2^2(\mu), \qquad
M_1(\mu) = {5\over 3}{M_0\over g_U^2}g_1^2(\mu),
\label{eqx}
\end{equation}
where we assume the gauge coupling unification at the scale $M_U$ 
and define a value of the gauge coupling at $M_U$ as $g_{_U}$. 
It is not unnatural to consider the gauge coupling of U(1)$_X$ and 
its gaugino mass to be the same as the ones of U(1)$_Y$ \footnote{In the 
superstring context the freedom of an Abelian Kac-Moody level can make 
it possible.}.
If we adopt this simplified possibility, we can find that $m_{1,3}$ and
$\sin^22\theta$ can be written by 
using only $q_\alpha$ and ${g_U^2\over \vert M_0\vert}u^2$ as
\begin{eqnarray}
&&m_{1,3}={3\over 5}\left(-(2+2q_1^2+q_2^2) \pm
\sqrt{(2+2q_1^2+q_2^2)^2-{16\over 3}
(q_1-q_2)^2}\right){g_U^2u^2\over \vert M_0\vert}, \nonumber \\
&&\sin^22\theta={8(2+3q_1q_2)^2\over (2+6q_1^2-3q_2^2)^2 +8(2+3q_1q_2)^2},
\label{eqy}
\end{eqnarray}
where we assume $M_0<0$.
The structure of the mass spectrum and the flavor mixing
is controlled by the U(1)$_X$-charge. The gaugino mass $M_0$ and the
VEV $u$ of sneutrinos are relevant to the 
mass eigenvalues only in the form of an overall factor 
${g_U^2\over \vert M_0\vert}u^2$.
In order to realize the value in the range of
$2\times 10^{-3}$~eV$^2~{^<_\sim}~\Delta m_{13}^2\simeq
\Delta m_{23}^2~{^<_\sim}~6\times 10^{-3}$~eV$^2$ which is
required by the atmospheric neutrino deficit, ${g_U^2\over \vert M_0\vert}u^2$ 
needs to be in the range of 0.003 eV~
${^<_\sim}~{g_U^2\over \vert M_0\vert}u^2~
{^<_\sim}~0.014$ eV.
If we take $\vert M_0\vert\sim 100$~TeV \footnote{
The gaugino mass is required to be larger than the usually considered
value. This comes from the $M_{Z_2}$ value imposed by the FCNC
constraint which has been discussed in the previous section.} 
and $g_{_U}\sim$0.72, for example,
it shows that the sneutrino VEV should be $u\sim 0.76$ - 1.6~MeV.
The remaining freedom which we can use to explain the solar neutrino
deficit is the U(1)$_X$-charge of ${\bf 5}^\ast_f$.  
\input epsf 
\begin{figure}[tb]
\begin{center}
\epsfxsize=7.6cm
\leavevmode
\epsfbox{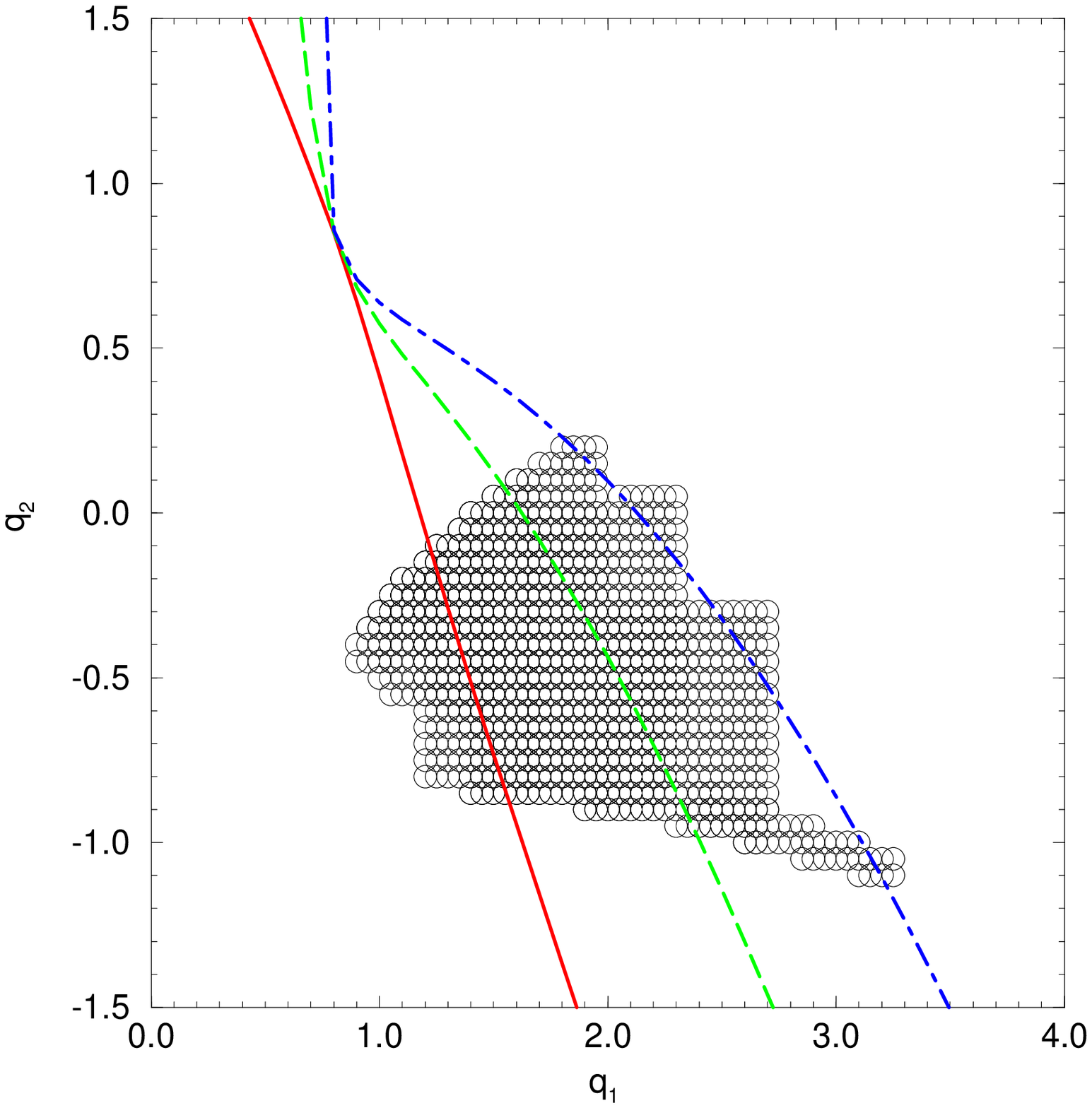}
\end{center}
\vspace*{-1cm}
{\footnotesize Fig. 1~~\  Scatter plots of the U(1)$_X$-charge
of neutrinos which can give the solution to the atmospheric and solar
neutrino problems. Solid, dashed and dot-dashed lines are the ones 
corresponding to $x=1.5,~3$ and 4.5, respectively.}
\end{figure}
In Fig.~1 we plot the value of the U(1)$_X$-charge to realize the LMA 
for the explanation of the solar 
neutrino deficit which requires $\cos\theta\sim 1$ as discussed above. 
Here we require $\cos\theta>0.98$ and also
$0.1\times 10^{-4}$~eV$^2~{^<_\sim}~\Delta m_{12}^2~{^<_\sim}~6\times
10^{-4}$~eV$^2$ to draw the figure. 
It shows that the reasonable value of the U(1)$_X$-charge
can realize the LMA. We should also note that the U(1)$_X$ charges obtained
here can satisfy the FCNC constraint given in (\ref{eqll}) coming from the
non-universal couplings of $Z^\prime$ as far as $\sin\xi<10^{-6}$ and 
$M_{Z_2}>100$~TeV are satisfied.
 
One of the unsolved important points is an origin of the small VEV 
$u$ of sneutrinos.
As mentioned in the previous part, it should be around $O(1)$~MeV
which is much smaller than the weak scale.
In the MSSM there are arguements on the lepton number violation 
due to the VEVs of sneutrinos in the vicinity of the weak scale \cite{rviol} 
and also some authors point out that the neutrino mass produced by 
them can be sufficiently small \cite{ratm}.
However, in our scenario we need much smaller VEVs of sneutrinos than
the weak scale. 
To consider such a possibility it is useful to note that
the small VEVs of sneutrinos could be obtained if there are
bi-linear $R$-parity violating terms $\epsilon L_\alpha H_2^1$
with a sufficiently small $\epsilon$. For example,
as such a candidate we may consider the non-renormalizable terms,
which are consistent with the U(1)$_F\times$U(1)$_X$ symmetry such as
\begin{equation}
{\bar NN\over M_{\rm pl}^2}{S_0\over M_{\rm pl}}S_{\epsilon_1}L_1H_2^1, \quad 
{\bar NN\over M_{\rm pl}^2}S_{\epsilon_2}L_2H_2^1,  \quad 
{\bar NN\over M_{\rm pl}^2}S_{\epsilon_3}L_3H_2^1,
\label{eqxx}
\end{equation}
where new SM singlet fields $S_{\epsilon_\alpha}$ are introduced for 
the U(1)$_X$-invariance. If an intermediate scale is induced through 
the $D$- and $F$-flat direction $\bar N=N$ of other singlet fields $N, 
~\bar N$ \cite{z00} and $S_{\epsilon_\alpha}$ acquires the VEV 
at the TeV scale, the appropriate $\epsilon$ term
might be obtained. However, their equality is not quaranteed in the present
example.
Once we find that $\epsilon$ terms could exist,    
we can check that the small $u$ is realized by minimizing the scalar
potential.
Under the assumption that $\langle H_1^{\ell}\rangle$ and $\langle
H_2^1\rangle$ can be treated as 
constants, the value of $u$ derived from the potential minimization 
is approximately expressed as
\begin{equation}
u \sim \epsilon{\mu \langle H_1^\ell\rangle+B_\epsilon\langle H_2^1\rangle 
\over m^2},
\end{equation}
where $B_\epsilon$ is a soft supersymmetry breaking parameter 
related to the $\epsilon L_\alpha H_2^1$ terms and $m^2$ is the soft 
scalar mass of sneutrinos, which is taken to be universal here. 
From this we find that the sufficiently small $u$ can be obtained as far 
as $\epsilon$ is small enough and the $\mu$-parameter, $B_\epsilon$ 
and $m^2$ take the values of the order of weak scale. 
We need to check whether these conditions are satisfied  
at the true vacuum by taking account of the radiative correction. 
Even if the VEVs of sneutrinos are not equal and eq.~(\ref{eqi}) is 
somehow modified, which may be expected in the case of (\ref{eqxx}),
there are two non-zero mass eigenvalues and the similar result obtained
above might be derived as far as the deviations from the equal $u$ are not so
large. The quanitative analysis on this aspect is also necessary to
know the viability of the model. Supersymmetry breaking scenario is
also important for the model. These issues are now under investigation. 
\vspace*{3mm}

\noindent
{\Large\bf 5.~Summary}

We have proposed the scenario for the origin of the mixings of
quarks and leptons in the supersymmetric model with an extra U(1)$_X$-symmetry.
The scenario is based on the usual Froggatt-Nielsen mechanism for the
quark and charged lepton sectors.
On the other hand, the mixing of neutrinos is considered to come from the
mixing among neutrinos and gauginos induced by the R-parity violation. 
In this model we could obtain two non-zero mass eigenvalues for
neutrinos at the tree level. 
The atmospheric and solar neutrino deficits can be simultaneously 
explained by the usual mass hierarchy scenario. 
In particular, the large mixing angle MSW solution for the solar
neutrino problem can be realized consistently with the small quark
mixings only by tuning the U(1)$_X$-charge of neutrinos. 
One of the important feature of the model is the value of $V^{(\rm
MNS)}_{e3}$ which can take a rather large value within the CHOOZ
constraint. 
The scenario gives an altenative possibility to the flavor mixing of
quarks and leptons as compared to the usual one. That is,
although the quarks and charged leptons have the origin of their mixings 
at the high energy region, the neutrinos may have it at the low energy region.
The most crucial unsolved problem is to clarify the R-parity violation and
the realization of the small sneutrino VEVs quantitatively. 
Further investigation of the problems seems to be necessary 
to see whether our model works well in a realistic way.
\vspace*{3mm}

\noindent
{\Large\bf Acknowledgement}\\ 
The Author thanks M.~Tanimoto for his useful comments.
This work is supported in part by the Grant-in-Aid for Scientific 
Research from the Ministry of Education, Science and Culture
(No.11640267).

\newpage

\noindent
{\Large\bf Appendix}

Non-trivial anomaly free conditions in the model are summarized as
\begin{equation}
\begin{array}{ccl}
{\rm SU(3)}^2{\rm U(1)}_X~&:  &
9\alpha +2q_1 +q_2 + x +y+z+w=0, \\
{\rm SU(2)}^2{\rm U(1)}_X~&: &
9\alpha +2q_1 +q_2 + p+q+r+s=0, \\
{\rm U(1)}_Y^2{\rm U(1)}_X~&: &
45\alpha + 5(2q_1 +q_2) + 2(x +y+z+w)+ 3(p+q+r+s)=0, \\
{\rm U(1)}_Y{\rm U(1)}_X^2~&: & 
p^2-q^2+r^2-s^2+y^2-x^2+w^2-z^2=0, \\
{\rm U(1)}_X^3~&: &
30\alpha^3+5(2q_1^3+q_2^3)+3(x^3+y^3+z^3+w^3)+ \\
& & \hspace*{3cm}+2(p^3+q^3+r^3+s^3)+\displaystyle{\sum_{i=1}^6}Q_i^3=0.\\
\end{array}
\label{eqzy}
\end{equation}
Combining eqs.~(\ref{eqzy}) except for the last one with 
eqs.~(\ref{eqc}) and (\ref{eqe}), we can express other parameters by
$q_1$, $q_2$ and $x$ as follows,
\begin{equation}
\begin{array}{lll}
\alpha=-{1\over 9}(2q_1+q_2), &p={2\over 9}(2q_1+q_2), & \\ 
q={1\over 9}(-7q_1+q_2), 
& r= {1\over 9}(q_1+5q_2),  & s= {1\over 9}(2q_1-8q_2), \\
y=-{1\over 3}(q_1-q_2)-x, & z=-{1\over 3}(q_1-q_2)+x, &
w={2\over 3}(q_1-q_2)-x, \\
Q_1={1\over 3}(q_1-q_2), & Q_2=-{2\over 3}(q_1-q_2), &
Q_3={1\over 3}(q_1-q_2),   \\
Q_4={2\over 3}(q_1-q_2), & Q_5=-{1\over 3}(q_1-q_2), &
Q_6=-{1\over 3}(q_1-q_2).   \\
\end{array}
\label{eqzz}
\end{equation} 
If we use eqs.~(\ref{eqzz}) in the last one of eqs.~(\ref{eqzy}) to solve it
numerically, we can plot the solutions in the $(q_1, q_2)$ plane for
each value of $x$. In Fig.~1 we show it for three values of $x$.   

\newpage


\begin{thebibliography}{99}
\bibitem{oscil1}Y.~Fukuda {\it et at.}, Phys. Rev. Lett. {\bf 81} (1998) 1562;
R.~Becker-Szendy {\it et al.}, IMB collaboration, Nucl. Phys. {\bf B38}
(proc. suppl.) (1995) 331; W.~W.~M.~Allison {\it et al.}, Soudan
collaboration, Phys. Lett. {\bf B391} (1997) 491; M.~Ambrosio {\it et 
al.}, Phys. Lett. {\bf B434} (1998) 451.

\bibitem{oscil2}B.~T.~Cleveland {\it et al.}, Ap. J. {\bf 496} (1998) 505;
K.~S.~Hirata {\it et al.}, Kamiokande collaboration, Phys. Rev. Lett. {\bf 77}
(1996) 1683; W.~Hampel {\it et al.}, GALLEX collaboration, Phys. Lett. {\bf
B388} (1996) 384; J.~N.~Abdurashitov {\it et al.}, SAGE collaboration, 
Phys. Rev. Lett. {\bf 77} (1996) 4708.

\bibitem{sk}Y.~Suzuki, Super-Kamiokande collaboration, 
Talk presented at {\it Neutrino 2000}, Sudbury, Canada. 

\bibitem{gut}S.~Bludman, N.~Hata, D.~Kennedy and P.~Langacker,
	Nucl. Phys. Proc. Supp. {\bf 31} (1993) 156; 
K.~S.~Babu and S.~M.~Barr, Phys. Lett. {\bf B381} (1996) 202;
J.~Sato and T.~Yanagida, Phys. Lett. {\bf B430} (1998) 127;
B.~Brahmachari and R.~N.~Mohapatra, Phys. Rev. {\bf D58} (1998) 015001;
C.~Albright, K.~S.~Babu, and S.~Barr, Phys. Rev. Lett. {\bf 81} (1998) 1167;
K.~S.~Babu, J.~C.~Pati and F.~Wilczek, Nucl. Phys. {\bf B566} (2000) 33; 
M.~Bando, T.~Kugo and K.~Yoshioka, Phys. Rev. Lett. {\bf 80} (1998)
	3004; 
K.~Oda, E.~Takasugi, M.~Tanaka and M.~Yoshimura, Phys. Rev. 
{\bf D59} (1999) 055001;
T.~Blazek, S.~Raby and K.~Tobe, Phys. Rev. {\bf D60} (1999) 113001;
Q.~Shafi and Z.~Tavartklidze, Phys. Lett. {\bf B487} (2000) 145;
K.~S.~Babu and S.~M.~Barr, Phys. Rev. Lett. {\bf 85} (2000) 1170. 

\bibitem{af}G.~Altarelli and F.~Feruglio, Phys. Lett. {\bf B451} (1999) 388.

\bibitem{fn}C.~Froggatt and H.~B.Nielsen, Phys. Lett. {\bf B147} (1979) 277.

\bibitem{fnap}P.~P.~Bin\'etruy, S.~Lavignac, S.~Petcov and P.~Ramond,
Nucl. Phys. {\bf B496} (1997) 3; 
P.~Bin\'etruy, N.~Irges, S.~Lavignac and P.~Ramond,
	Phys. Lett. {\bf B403} (1997) 38;
Y.~Grossman, Y.~Nir and Y.~Shadmi, JHEP {\bf 9810} (1998) 007;
Y.~Nir and Y.~Shadmi, JHEP {\bf 9905} (1999) 023. 

\bibitem{seesaw}M.~Gell-Mann, P.~Romond and R.~Slansky, in {\it
	Supergravity}, eds. P.~van Nieuwenhuizen and D.~Freedman
	(North-Holland, Amsterdam, 1979) p.315; T.~Yanagida, in {\it
	Proc. Workshop on Unified Theory and Baryon Number in the
	Universe}, eds. O.~Sawada and A.~Sugamoto (KEK, 1979).  

\bibitem{gaugino}D.~Suematsu, hep-ph/0012321 (to be published in
	Phys. Lett. {\bf B}).

\bibitem{rparity1}F.~de Compos, M.~A.~Garc\'a-Jare\~no, A.~S.~Joshipura, 
J.~Rosiek and J.~W.~F.~Valle, Nucl. Phys. {\bf B451} (1995) 3;
T.~Banks, Y.~Grossman, E.~Nardi and Y.~Nir, Phys. Rev. {\bf D52} (1995)
5319; A.~S.~Joshipura and M.~Nowakowshi, Phys. Rev. {\bf D51}
(1995) 2421.

\bibitem{rparity2}M.~A.~D\'iaz, J.~C.~Rom\~ao and J.~W.~F.~Valle,
Nucl. Phys. {\bf B524} (1998) 23; J.~W.~F.~Valle, 
in {\it Pro. 6th International Symposium on Particles, Strings and 
Cosmology} (PASCOS 98), Boston 1998, p502 and references therein.

\bibitem{z00}D.~Suematsu and Y.~Yamagishi, Int. J. Mod. Phys. {\bf A10} 
(1995) 4521; M.~Cveti$\check{\rm c}$ and P.~Langacker, 
Phys. Rev. {\bf D54} (1996),
Mod. Phys. Lett. {\bf A11} (1996) 1247; D.~Suematsu, Phys. Rev. {\bf
	D59} (1999) 055017. 

\bibitem{sue2}E.~Nardi, Phys. Rev. {\bf D48} (1993) 3277;
	Phys. Rev. {\bf D49} (1994) 4394; E.~Nardi and T.~Rizzo,
	Phys. Rev. {\bf D50} (1994) 203;
D.~Suematsu, Prog. Theor. Phys. {\bf 96} (1996) 611.

\bibitem{gm}G.~F.~Giudice and A.~Masiero, Phys. Lett. {\bf B206} (1988) 480.
 
\bibitem{fcnc}P.~Langacker and M.~Plumacher, Phys. Rev. {\bf D62} (2000) 
	013006.

\bibitem{sue1}D.~Suematsu, Mod. Phys. Lett. {\bf A12} (1997) 1709;
	Phys. Lett. {\bf B416} (1998) 108; Phys. Rev. {\bf D57} (1998) 1738.

\bibitem{bimaxim}V.~Barger, S.~Pakvasa, T.~J.~Weiler and K.~Whisnant,
	Phys. Lett. {\bf B437} (1998) 107; S.~Davidson and S.~F.~King,
	Phys. Lett. {\bf B445} (1998) 191; R.~N.~Mohapatra and
	S.~Nussinov, Phys. Rev. {\bf D60} (1999) 013002. 

\bibitem{bimaxim2}I.~Stancu and D.~V.~Ahluwalia, Phys. Lett. {\bf B460}
	(1999) 431.

\bibitem{bimaxim3}For the recent models with the large mixing, 
see for example, 
J.~M.~Mira, E.~Nardi, D.~A.~Restrepo and
	J.~W.~F.~Valle, Phys. Lett. {\bf 492} (2000) 81;
A. de Gouvea and J.~W.~F.~Valle, Phys. Lett. {\bf B501} (2001) 115;
P.~H.~Chankowski, A.~Ioannisian, S.~Pokorski and J.~W.~F.~Valle,
	hep-ph/0011150; 
Q.~Shafi and Z.~Tavartkiladze, hep-ph/0101350; 
S.~F.~King and M.~Oliveira, hep-ph/0009287.

\bibitem{chooz}C.~Bemporad, CHOOZ collaboration, Nucl. Phys. B
	(Proc. Suppl.) {\bf 77} (1999) 159.

\bibitem{ue3}G.~Fogli, E.~Lisi, A.~Marrone and G.~Scioscia,
	Phys. Rev. {\bf D59} (1999) 033001; S.~Bilenky, G.~Giunti and
	W.~Grimus, hep-ph/9809368.

\bibitem{bb0}S.~M.~Bilenky, C.~Giunti, C.~W.~Kim and S.~T.~Petcov,
	Phys. Rev. {\bf D54} (1996) 4432; S.~M.~Bilenky, C.~Giunti,
	W.~Grimus, B.~Kayser and  S.~T.~Petcov, Phys. Lett. {\bf B465}
	(1999) 193; H.~V.~Klapdor-Kleingrothaus, H.~P\"as and A.~Yu.~Smirnov,
	hep-ph/0003219. 

\bibitem{zee}A.~Zee, Phys. Lett. {\bf 93B} (1980) 389; Phys. Lett. {\bf
	161B} (1985) 141.

\bibitem{rviol}A.~Masiero and J.~W.~F.~Valle, Phys. Lett. {\bf B251}
	(1990) 273; J.~C.~Rom\~ao, C.~A.~Santos and J.~W.~F.~Valle,
	Phys. Lett. {\bf B288} (1992) 311;
J.~C.~Rom\~ao, A.~Ioannisian and J.~W.~F.~Valle,
	Phys. Rev. {\bf D55} (1997) 427.

\bibitem{ratm}J.~C.~Rom\~ao, M.~A.D\'iaz, M.~Hirsch, W.~Porod 
and J.~W.~F.~Valle, Phys. Rev. {\bf D61} (2000) 071703;
M.~Hirsch, M.~A.D\'iaz, W.~Porod, J.~C.~Rom\~ao and J.~W.~F.~Valle,
Phys. Rev. {\bf D62} (2000) 113008. 

\end{thebibliography}
\end{document}